\documentclass[aps,prl,superscriptaddress,notitlepage,nopacs,amsmath,amstex,amssymb,citeautoscript,longbibliography,floatfix,letter,twocolumn]{revtex4-2}
\usepackage{graphicx}
\usepackage{subfigure}
\usepackage{adjustbox}
\usepackage{bm}
\usepackage{ulem}
\usepackage{color}
\usepackage{braket}
\usepackage{standalone}
\usepackage{multirow}
\usepackage{tikz}
\usepackage{mathrsfs}
\usepackage{dsfont}
\usepackage{comment}
\usepackage{amsmath, amsthm}

\usepackage{dcolumn}% Align table columns on decimal point

\begin{document}
	\title{Spin-Orbit Structure and Helicity anomaly in Relativistic Electron Vortex Beams}
	
	\author{Zhongze Guo}
	\email{guozhongze007@gmail.com}
	\affiliation{Department of Physics and Institute of Theoretical Physics, University of Science and Technology Beijing, Beijing 100083, China}
	
	\author{Bei Xu}
	\email{xubei0903@163.com}
	\affiliation{Institute for Advanced Study, Tsinghua University, Beijing 100084, China}	
	
	\author{Qiang Gu}
	\email[Corresponding author: ] {qgu@ustb.edu.cn}
	\affiliation{Department of Physics and Institute of Theoretical Physics, University of Science and Technology Beijing, Beijing 100083, China}

	\begin{abstract}
		The relativistic electron vortex beam (REVB) has attracted increasing attention due to its nontrivial spin-orbit structure recently. As relativistic electrons are governed by the Dirac equation, exact solutions to this equation provide the most reliable starting point for understanding angular momentum characteristics of REVBs. In this work, a set of exact eigensolutions of the Dirac equation are derived in a complex cylindrical coordinate system using a generalized series expansion method. We demonstrate that the eigenstate carries net angular momentum with the vortex charge being the quantum number of the total angular momentum along the propagation direction and deduce the explicit expression for the intrinsic spin-orbit coupling strength. Furthermore, we show that helicity, which exhibits anomaly in the vortex state, can serve as a practical characterizing quantity for the REVB. This work lays a theoretical foundation for further exploration of REVBs in both theory and experiment.
	\end{abstract}
	
	\maketitle
	
	\textit{Introduction -}
	The concept of a vortex electron—an electron whose wavefunction possesses a helical phase structure—was first introduced in 2007 by Bliokh {\it et al.} \cite{electron07} and experimentally realized in 2010 \cite{electron10,electron10b,electron11}. Since then, electron vortex beams have opened up new avenues in electron microscopy, atomic-scale manipulation, and fundamental physics \cite{elecron17pr,electron17rmp}. The central feature of the vortex beam is that it carries certain orbital angular momentum. More recently, relativistic electron vortex beams (REVBs) have attracted increasing interest \cite{relectron11,relectron17b,relectron17,zhang18,zhang19}, owing to two-fold motivations. First, experimental advancements make it possible to generate electron beams with high energies (100-300keV) at which relativistic effects need to be incorporated. Secondly, the spin and orbital angular momentum of electrons are not separately conserved in the relativistic case and thus the description of the angular momentum based on the Schr{\"o}dinger equation becomes inadequate. Therefore, a full description of angular momentum properties of REVBs should start with solutions of the Dirac equation, the fundamental equation for the relativistic electrons. 
	
	Current theory in this regard is still under controversy. So far, the most representative theoretical works on this topic include three papers published in PRL \cite{relectron11,relectron17b,relectron17}. Bliokh {\it et al.} first demonstrated that electron vortexes can be formed in the relativistic case as does in the low energy case, and REVBs exhibit nontrivial vortex-dependent properties in comparison to the low energy vortex beams \cite{relectron11}. They suggested the vortex charge is the total angular momentum in the $z$-direction and predicted that the spin and orbital angular momentum is intrinsic coupled. To further investigate properties of REVBs, Bialynicki-Birula duo \cite{relectron17b} and Barnett \cite{relectron17} constructed two vortex solutions of the Dirac equation using different approximation methods. However they arrived at opposite conclusions. Barnett's results basically agree with Bliokh {\it et al.}'s suggestion \cite{relectron17}, but lake of information of the spin-orbit coupling (SOC) \cite{note1}. Nevertheless, Bialynicki-Birula duo supported the existence of SOC, but argued that the beam unlikely carries net angular momentum since the orbital part of the angular momentum might be canceled by the spin part \cite{relectron17b}. These conflicts have sparked a series of debates on the definitions and understandings of vorticity in fluid mechanics and angular momentum in quantum mechanics \cite{comment17a,replies17a,viewpoint2017new}, leaving it an open question whether relativistic vortex electrons can indeed exhibit true vortex behavior as claimed in experiments \cite{viewpoint2017new}. 
	
	We notice that above theoretical frameworks for describing REVBs is based on constructed wavefunctions in cylindrical coordinates by superimposing eigenstates of the Dirac equation in Cartesian coordinates through some approximations. In order to verify current consensus and controversy, it is necessary to perform a comprehensive study based on the rigorous solution of the Dirac equation. It is just the central goal of our present work. Although several exact solutions of the Dirac equation in cylindrical coordinates have been known for decades, they are typically derived in context of either the quark production in a quantum backreaction problem \cite{PhysRevD}, or the motion of electrons and neutrinos in a background gravitational field \cite{RevModPhys.29.465}. These solutions do not care about the information related to angular momentum (the wave function appears in the form of summation of angular momentum) and thus are not convenient to be applied directly to REVBs \cite{elecron17pr}. 
	
	In this letter, we employ a generalized series expansion technique to construct explicit vortex eigensolutions of the Dirac equation. These solutions are rigorous eigenfunctions of the total angular momentum in the $z$-direction with non-zero eigenvalues, and thus confirms the existence of relativistic electron vortex state straightforwardly. Moreover, we propose that helicity can serve as an alternative quantity to characterize REVBs. The helicity exhibits anomalous behaviors, owing to the broken of translational invariance perpendicular to the $z$-direction. 
	
	\textit{The Dirac equation in cylindrical coordinates -}
	The relativistic electron is described by the Dirac equation,
	\begin{equation}
		\hat{H}|\psi\rangle=(-ic\hbar \vec{\alpha}\cdot \vec{\nabla}+\beta mc^{2})|\psi\rangle=E|\psi\rangle ,
		\label{ham}
	\end{equation}
	with $m$ being the rest mass of the electron and $E$ the eigenenergy. $|\psi\rangle=(\psi_1,\psi_2,\psi_3,\psi_4)^{T}$ is a four-component spinor wavefunction and the Dirac matrices take the standard form in the Dirac-Pauli representation,
	\[
	\beta = \begin{pmatrix} I & 0 \\ 0 & -I \end{pmatrix}, \quad
	\vec{\alpha} = \begin{pmatrix} 0 & \vec{\sigma} \\ \vec{\sigma} & 0 \end{pmatrix},
	\]
	where $\vec{\sigma}=(\sigma^x,\sigma^y,\sigma^z)$ are the Pauli matrices and $I$ is the $2\times2$ identity matrix. 
	To describe vortex states, it is necessary to express the Dirac equation in cylindrical coordinates $(r,\theta, z)$. We note that the Dirac equation in the Dirac-Pauli representation can be naturally transformed form the Cartesian coordinate system into the complex cylindrical coordinates, 
	\begin{align}
		\frac{E - mc^2}{ic} \psi_{1} + e^{-i\theta}\left(\hbar\frac{\partial}{\partial r} - i\frac{\hbar}{r}\frac{\partial}{\partial \theta}\right)\psi_{4} 
		  + \hbar \frac{\partial}{\partial z}\psi_{3} &= 0, \nonumber\\ 
		\frac{E - mc^2}{ic} \psi_{2} + e^{-i\theta}\left(\hbar\frac{\partial}{\partial r} + i\frac{\hbar}{r}\frac{\partial}{\partial \theta}\right)\psi_{3} 
		  - \hbar \frac{\partial}{\partial z}\psi_{4} &= 0, \nonumber\\ 
		\frac{E + mc^2}{ic} \psi_{3} + e^{-i\theta}\left(\hbar\frac{\partial}{\partial r} - i\frac{\hbar}{r}\frac{\partial}{\partial \theta}\right)\psi_{2} 
		  + \hbar \frac{\partial}{\partial z}\psi_{1} &= 0, \nonumber\\
		\frac{E + mc^2}{ic} \psi_{4} + e^{-i\theta}\left(\hbar\frac{\partial}{\partial r} - i\frac{\hbar}{r}\frac{\partial}{\partial \theta}\right)\psi_{1} 
		  - \hbar \frac{\partial}{\partial z}\psi_{2} &= 0.
	\label{eq}
	\end{align}
	Mathematical details are given in the Supplementary Material (SI).
	
	\textit{The generalized series expansion -}
	The general approach to solve the Dirac equation in cylindrical coordinates is to convert the four coupled first-order differential equations into two second-order differential equations through methods such as Foldy-Wouthuysen (FW) transformation under a low-energy approximation \cite{zhang18,zhang19}, or through applying the charge conjugation operator to the covariant generalization of the Dirac equation \cite{booklifi1982}. However, these transformations require the presence of an external field and therefore are usually invalid for the free-electron Dirac equation. Moreover, such treatments lead to the loss of intrinsic spin entanglement information which is closely related to the SOC or helicity. 
	
	We employ a generalized power series expansion method to solve the Dirac equation analytically. Each component of the spinor wavefunction is expanded in terms of cylindrical harmonics, $\psi_s(r,\theta, z) = R_{s}(r)e^{in_{s}\theta}e^{i p_{z}z/\hbar}$, incorporating azimuthal dependence via $e^{in_s \theta}$, where $n_s$ is the azimuthal quantum number. The radial parts are expanded in power series of $r$ as $R_s(r) = r^\alpha \sum_{k=0}^{\infty} C_k^s r^k$, and the longitudinal part is represented using a plane wave $e^{ik_z z}$ where $k_z=p_{z}/\hbar$ is the momentum quantum number along $z$ direction. The radial functions $R_{s}(r)$ are determined by solving the resulted recurrence relations as shown in Supplemental Material (SII). We prove that the power series solution for the radial function corresponds to Bessel functions, $J_{n}(\kappa r)$, with $p_\kappa=\hbar\kappa=\sqrt{E^2/c^2-m_0^2c^2-p_z^2}$. Thus, the spinor wave function is given by 
	\begin{eqnarray}
		\begin{aligned}
			&|\psi_{n,\kappa,k_{z}}\rangle \\ 
			&= e^{\frac{i}{\hbar}p_z z}e^{i n\theta}
			\begin{bmatrix}
				J_{n}(\kappa r)    \\
				\frac{-i}{\hbar \kappa} \left( p_z-\frac{E+mc^2}{c\lambda} \right) J_{n+1}(\kappa r) e^{i \theta} \\
				\frac{1}{\lambda}J_{n}(\kappa r) \\
				\frac{-i}{\hbar \kappa} \left( \frac{p_z}{\lambda}-\frac{E-mc^2}{c} \right) J_{n+1}(\kappa r) e^{i \theta}
			\end{bmatrix} ,
		\end{aligned}  
		\label{wavefunction0}
	\end{eqnarray}
    which is just the eigen solution of Eq. (\ref{ham}). However, this solution is incomplete since the parameter $\lambda$ is not yet determined so far. It is necessary to introduce an additional conserved quantity to fix the value of $\lambda$ and identify each separate eigenstate definitely. 
	
	\textit{The auxiliary conserved quantity - }
	The auxiliary conserved quantity $\hat{K}$ used to determine the value of $\lambda$ is obtained through the algebraic method of separation of variables applied to the Dirac equation by a similarity transformation to diagonalize the Vierbein matrices \cite{Villalba1989,PhysRevD}. As shown in Supplemental Material (SIII), this new conserved quantity $\hat{K}$ is derived as 
	\begin{widetext}
		\begin{eqnarray}
			\hat{K} |\tilde{\Phi}\rangle = \left[ \gamma^{1}\gamma^{0}\gamma^{3}\cos \theta \frac{\partial}{\partial r}
			-\gamma^{1}\gamma^{0}\gamma^{3}\frac{\sin \theta}{r}\frac{\partial}{\partial \theta}
			+\gamma^{2}\gamma^{0}\gamma^{3}\sin \theta \frac{\partial}{\partial r}
			+\gamma^{2}\gamma^{0}\gamma^{3}\frac{\cos \theta}{r}\frac{\partial}{\partial \theta} \right] |\tilde{\Phi}\rangle
			=\lambda' |\tilde{\Phi}\rangle.
		\end{eqnarray}
	\end{widetext}
    The operator $\hat{K}$ commutes with other relevant operators, including $\hat{H}$, $\hat{p}_z$ and $\hat{J}_z$ and the wave function obeys the equation $|\psi\rangle=\gamma^3\gamma^0|\tilde{\Phi}\rangle$. Obviously, 
    \begin{eqnarray}
    	\hat{K} |{\psi}\rangle =\lambda' |{\psi}\rangle
    \end{eqnarray}
    and $\hat{K}^2 |{\psi}\rangle =p_{\kappa}^2 |{\psi}\rangle$, with corresponding eigenvalues $\lambda'=\pm p_{\kappa}$. For $\lambda'=p_{\kappa}$, we get $\lambda=\frac{p_z+ip_{\kappa}}{E-mc^2}$ and the normalized wave function reads
	\begin{eqnarray}
		\begin{aligned}
			|\psi_{n,\kappa,k_z,\tilde{\lambda}}\rangle
			&=N e^{\frac{i}{\hbar}p_z z}e^{i n\theta}
			\begin{bmatrix}
				J_{n}(\kappa r)    \\
				J_{n+1}(\kappa r) e^{i \theta} \\
				\frac{k_z-i\kappa}{E+m}J_{n}(\kappa r) \\
				-\frac{k_z-i\kappa}{E+m}J_{n+1}(\kappa r) e^{i \theta}   
			\end{bmatrix} .
		\end{aligned}
		\label{wavefunction1}
	\end{eqnarray}	
	For $\lambda'=-p_{\kappa}$, $\lambda=\frac{p_z-ip_{\kappa}}{E-mc^2}$ and 
	\begin{eqnarray}
		\begin{aligned}
			|\psi_{n,\kappa,k_z,\tilde{\lambda}}\rangle
			&=N e^{\frac{i}{\hbar}p_z z}e^{i n\theta}
			\begin{bmatrix}
				\frac{k_z-i\kappa}{E+m}J_{n}(\kappa r)    \\
				\frac{k_z-i\kappa}{E+m}J_{n+1}(\kappa r) e^{i \theta} \\
				J_{n}(\kappa r) \\
				-J_{n+1}(\kappa r) e^{i \theta}   
			\end{bmatrix} .
		\end{aligned}
		\label{wavefunction2}
	\end{eqnarray}	
	Here $N=\sqrt{\frac{E+mc^2}{(4\pi E D I_1)}}$ is the normalization coefficient, where $D$ is the beam length, arising from the $z$ integration for a beam extending between $-\frac{D}{2}$ and $\frac{D}{2}$, and $I_1$ is defined as
	\begin{eqnarray}
		I_1=\int_0^{r_1}\left[J_{n}^2(\kappa r)+J_{n+1}^2(\kappa r)\right]rdr ,
	\end{eqnarray}	
	where the upper limit of the integration $r_1$ could be infinity. During performing numerical calculation, a cutoff of $r_1$ is needed to avoid divergence of the integration \cite{Lloyd2012}. For convenience, the effective integral interval can be assumed within the first zero of the corresponding Bessel function at $\alpha$, i.e., $r_1=\frac{\alpha}{\kappa}$. 
	
	\textit{The vortex and spin-orbit coupling -}
	Eqs. (\ref{wavefunction1}) and (\ref{wavefunction2}) compose the set of eigenfunctions of the positive energy solution of the Dirac equation, which serve as the basis for discussing properties of REVBs. They are not only common eigenfunctions of the Hamiltonian $\hat{H}$, the longitudinal momentum $\hat{p}_z$, and the auxiliary operator $\hat{K}$, but also of the total angler momentum in the z-direction $\hat{J}_z$. Taking the wavefunction in Eq. (\ref{wavefunction1}) for example,   
	\begin{equation}
		\hat{J}_z|\psi\rangle = \left(\hat{ L}_z+\hat{S}_z\right)|\psi \rangle = \left(n\hbar+\frac{1}{2}\hbar\right) |\psi\rangle ,
		\label{jz}
	\end{equation}
	where $\hat{S}_z=\frac{\hbar}{2}\hat{\Sigma}_z$ and the matrix $\hat{\Sigma}_z = \begin{pmatrix} \sigma^z & 0 \\ 0 & \sigma^z \end{pmatrix}$. This result straightforwardly confirms the existence of electron vortexes in the relativistic regime \cite{relectron11,relectron17}.  
	
    Though $\psi_{n,\kappa,k_z}$ is the eigenfunction of $\hat{J}_z$, but not the eigenstate either of the orbital angular momentum operator $\hat{L}_{z}$ or the spin angular momentum operator $\hat{S}_{z}$. As stated in Ref. \cite{relectron11}, this result implies the existence of intrinsic SOC. In the following, we show that the SOC can be demonstrated more clearly by calculating expectation values of $\hat{L}_{z}$ and $\hat{S}_{z}$ in a given eigenstate. First, let's rewrite the eigen wave function as		
	\begin{eqnarray}
		|\psi_{n,\kappa,k_z}\rangle = e^{\frac{i}{\hbar} p_z z} e^{in\theta} |\varphi_n^{\kappa}\rangle ,
		\label{psi2}
	\end{eqnarray}
	where $e^{\frac{i}{\hbar} p_z z}$ is the free plane wave factor in the $z$-direction, $\varphi_n=e^{in\theta}$ denotes the orbital angular momentum part with $n$ the quantum number of the orbital angular momentum, $|\varphi_n^{\kappa}\rangle$ represents the Dirac spinor. Then the expectation values are given by
	\begin{eqnarray}
		&&\langle \hat{L}_z\rangle=n\hbar+ \frac{\langle \varphi_n^{\kappa }|(-i\hbar\frac{\partial}{\partial \theta})|\varphi_n^{\kappa}\rangle }{\langle\varphi_n^{\kappa}|\varphi_n^{\kappa}\rangle}=\left(n+\Delta_n\right)\hbar , \label{lbar}\\
		&& \langle \hat{S}_z\rangle=\frac{\langle \varphi_n^{\kappa }|(\frac{\hbar}{2}\hat{\Sigma}_z)|\varphi_n^{\kappa}\rangle }{\langle\varphi_n^{\kappa}|\varphi_n^{\kappa}\rangle} = \left(\frac{1}{2}-\Delta_n \right)\hbar ,  \label{sbar}
	\end{eqnarray} 
	where 
	\begin{eqnarray}
		\Delta_n = \frac{\langle \varphi_n^{\kappa }|(-i\frac{\partial}{\partial \theta})|\varphi_n^{\kappa}\rangle} {\langle\varphi_n^{\kappa}|\varphi_n^{\kappa}\rangle} = \frac{1}{I_1}\int_0^{r_1}J_{n+1}^2(\kappa r)rdr .
	\label{Delta}
	\end{eqnarray} 	
	Formally, Eqs. (\ref{lbar}) and (\ref{sbar}) are consistent with the Eq. (17) in Ref. \cite{relectron11}, with $\Delta_n$ characterizing the strength of SOC. Not only that, the present study gives an explicit expression of $\Delta_n$ in Eq. (\ref{Delta}), which indicates that $\Delta_n$ is directly related to vortex charge $n$. It is easy to get that $\Delta_n$ decreases with $n$ by performing the integration numerically. 
	
	More helpful information can be drawn from Eq. (\ref{Delta}). Note that only $|\varphi_n^{\kappa}\rangle$ appears in the expression of $\Delta_n$, while the other two part of wavefunction in Eq. (\ref{psi2}), $e^{\frac{i}{\hbar} p_z z}$ and $e^{in\theta}$, are not involved. That is to say, the SOC arises basically from the Dirac spinor. One side, this signifies that the SOC is naturally a kind of relativistic effect. On the other hand, it implies that this effect occurs at a very small length-scale: the sub-Compton wavelength, rather than the sub-de Broglie wavelength \cite{FW1950}. The SOC in relativistic electrons is fundamentally different from that in the vortex light. In the latter case, the effective scale of SOC emerges at sub-wavelength dimensions, making it accessible through optical experiments \cite{Bliokh2015}. 
	
	An alternative way has been proposed to characterize the vortex feature of relativistic electrons via the vorticity of the probability current \cite{relectron17b}. However, the applicability of the concept of vorticity from fluid mechanics to quantum mechanics remains a subject of debate \cite{comment17a,replies17a}. Moreover, it is worth questioning to calculate the vorticity of the probability current through the Gordon decomposition whose validity  relies on the assumption that linear momentum is a good quantum number—an assumption that is clearly violated in vortex states.
		
	\textit{The vortex and helicity -}
	The helicity formalism has been extensively developed to describe scattering processes and decays of particles in relativistic quantum theory \cite{Bjorken1964}, 
	as it can be used conveniently to obtain angular distributions in final states. The reason why helicity is suitable for handling relativistic problems lies in it commutes with the free-particle Hamiltonian and remains invariant under both spatial rotations and Lorentz boosts along the direction of momentum.
	
	In contrast to conventional plane-wave states, vortex electron states are not eigenstates of total linear momentum. They retain translationally invariant just along the propagation axis (typically taken as z-axis). And their transverse momentum forms a conical distribution-a hallmark of Bessel beams solution, as we have proven above. This key structural difference implies that the expectation value of helicity may differ between plane-wave and vortex electron states. Thus, helicity could serve as a distinguishing observable for probing the internal dynamics and symmetry properties of vortex electrons.
	
	The helicity operator is defined as
	\begin{equation}
		\hat{\Sigma} \cdot \hat{p}=
		\begin{bmatrix}
			\vec{\sigma} \cdot \vec{p}&0\\
			0 &\vec{\sigma} \cdot \vec{p}
		\end{bmatrix} .
	\end{equation}
	And in the cylindrical coordinate, the component
	\begin{equation}
		\vec{\sigma} \cdot \vec{p}=
		\begin{bmatrix}
			-i\hbar \frac{\partial}{\partial z} & e^{-i\theta}\hbar( \frac{\partial}{\partial r}-\frac{i}{r} \frac{\partial}{\partial \theta} )  \\
			-e^{i\theta}\hbar( \frac{\partial}{\partial r}+\frac{i}{r} \frac{\partial}{\partial \theta} ) & i\hbar \frac{\partial}{\partial z} 
		\end{bmatrix},
	\end{equation}
	
	The conventional plane-wave states of relativistic electron are eigenstates of the helicity operator with corresponding eigenvalue $\pm p$, indicating the parallel or anti-parallel alignment of the spin and momentum. In contrast, vortex electron states given above are not the eigenstates of helicity. Let's  consider the expectation value of helicity,
	\begin{equation}
		\langle \psi|\hat{\Sigma} \cdot \hat{p}|\psi\rangle 
		=\frac{(p_z-i \frac{1}{\gamma}p_\kappa) }{I_1}\int_0^{r_1}\left[J_{n}^2(\kappa r)-J_{n+1}^2(\kappa r)\right]rdr . \label{helicity}
	\end{equation}
	where $\frac{1}{\gamma}=\frac{mc^2}{E}=\sqrt{1-(\frac{v}{c})^2}$, $\gamma$ is the Lorentz factor. 
	The expectation value is no longer a real number, and the real and imaginary part correspond to contributions from the longitudinal component of the helicity ($\langle \hat{\Sigma}_z \hat{p}_z\rangle$) and the transverse one, respectively. The imaginary part indicates that the transverse component of the helicity becomes not well-defined and thus is not measurable in the vortex state. This seems anomalous, but actually a natural result of symmetry-broken of translational variance in the direction perpendicular to the propagation axis. Meanwhile, $\langle \hat{\Sigma}_z \hat{p}_z\rangle$ remains real, owing to the persistence of translational symmetry in $z$-direction. Therefore, it suggests that the helicity is still an observable. However, the expectation value is not equal to either the $z$ component of the momentum $p_z$ or total momentum value $p$. Its value increase with the topological charge $n$ of the vortex state, as Eq. (\ref{helicity}) denotes. Thus $n$ can determined by detecting $\langle \hat{\Sigma}_z \hat{p}_z\rangle$. At present, each experimental technique used to measure the angular momentum of vortex electrons has its own limitations \cite{elecron17pr}. As an alternative quantity for characterizing vortex properties of REVBs, helicity could be employed as a more applicable tool to distinguish vortex electron beams from conventional (non-vortex) electron beams in experiments.
	
	In prior studies, it was typically assumed that all plane waves share the same polarization amplitude. While this greatly simplifies the calculations, it makes it impossible to accurately evaluate the associated helicity \cite{relectron11,relectron17,relectron17b}.
	
	\textit{Concluding remarks -}
	In conclusion, we derive a set of exact eigensoutions of the Dirac equation which can be conveniently applied to describing relativistic electron vortexes and  preform a comprehensive theoretical study on vortex properties of the relativistic electron vortex beams on this basis. Such exact vortex eigensoutions provide a reliable starting point to verify the consensus and controversy of current theoretical frameworks on the REVB. As a result, we have straightforwardly confirmed that the relativistic electron beam can indeed carry net angular momentum with the vortex charge characterized by the quantum number of total angular momentum in the propagating direction. Moreover, we demonstrate that the spin and orbital angular momentum are intrinsically coupled as predicted previously and obtain a explicit expression for the coupling strength. To go further, our results reveal that the expectation value of helicity provides a meaningful characterization of the vortex structure of the REVB. 
	
	Q.G. acknowledges financial support from the National Natural Science Foundation of China through Grant No. 1874083.

	\bibliography{cyreference}

\begin{thebibliography}{24}
\bibitem [{\citenamefont {{The FW transformation is unitary. In the FW
  representation,the spin and position operators correspond to the mean spin
  angular momentum and the mean position, respectively. Consequently, even
  after applying the inverse FW transformation}}()}]{note1}%
  \BibitemOpen
  \bibfield  {author} {\bibinfo {author} {\bibnamefont {{The FW transformation
  is unitary. In the FW representation,the spin and position operators
  correspond to the mean spin angular momentum and the mean position,
  respectively. Consequently, even after applying the inverse FW
  transformation}}},\ }\bibfield  {title} {\bibinfo {title} {the associated
  observables—spin and position—remain averaged over the order of the
  compton wavelength. since the spin operator commutes with the hamiltonian},\
  }\href@noop {} {\ }\bibinfo {note} {Intrinsic spin–orbit coupling (SOC) is
  effectively lost.}\BibitemShut {Stop}%
\end{thebibliography}

\newpage 
\cleardoublepage

\onecolumngrid
	\begin{center}
		\textbf{\large Supplemental Material for "Spin-Orbit Structure and Helicity anomaly in Relativistic Electron Vortex Beams"}\\[5pt]
		
		\begin{center}
			{\small Zhongze Guo$^{1}$, Bei Xu$^{2}$ and Qiang Gu$^{1}$}  
		\end{center}
		
		\begin{center}
			{\sl \footnotesize
				$^{1}$Department of Physics and Institute of Theoretical Physics, University of Science and Technology Beijing, Beijing 100083, China
				
				$^{2}$Institute for Advanced Study, Tsinghua University, Beijing 100084, China
			}
		\end{center}

		\begin{quote}
			{\small This supplemental material contains some technical details. In Sec.~\ref{sec: Coordinate}, we show that the Dirac equation can be naturally transformed from the Cartesian coordinate system to the complex cylindrical coordinates. In Sec.~\ref{sec: Expansion}, we reproduce exact solutions of the Dirac equation via the series expansion method. Sec.~\ref{sec: Derivation} presents mathematical details in derivation of the conserved quantity $\hat{K}$.  
			}
		\end{quote}
	\end{center}

	\vspace*{0.4cm}
	
	\setcounter{equation}{0}
	\setcounter{figure}{0}
	\setcounter{table}{0}
	\setcounter{page}{1}
	\setcounter{section}{0}
	\makeatletter
	\renewcommand{\theequation}{S\arabic{equation}}
	\renewcommand{\thefigure}{S\arabic{figure}}
	\renewcommand{\thesection}{S\Roman{section}}
	\renewcommand{\thepage}{\arabic{page}}
	\renewcommand{\thetable}{S\arabic{table}}
	
	\section{The Coordinate Transformation}
	\label{sec: Coordinate}
	
	In the Cartesian coordinate system, the Dirac equation can be extended to a set of four coupled first-order partial differential equations,
	\begin{align}
		i \frac{\partial \psi_1}{\partial t} &= -i \left( \frac{\partial \psi_4}{\partial x} + i \frac{\partial \psi_4}{\partial y} + \frac{\partial \psi_3}{\partial z} \right) + mc^2 \psi_1 \\
		i \frac{\partial \psi_2}{\partial t} &= -i \left( \frac{\partial \psi_3}{\partial x} - i \frac{\partial \psi_3}{\partial y} - \frac{\partial \psi_4}{\partial z} \right) + mc^2 \psi_2 \\
		i \frac{\partial \psi_3}{\partial t} &= -i \left( \frac{\partial \psi_2}{\partial x} + i \frac{\partial \psi_2}{\partial y} + \frac{\partial \psi_1}{\partial z} \right) - mc^2 \psi_3 \\
		i \frac{\partial \psi_4}{\partial t} &= -i \left( \frac{\partial \psi_1}{\partial x} - i \frac{\partial \psi_1}{\partial y} - \frac{\partial \psi_2}{\partial z} \right) - mc^2 \psi_4
	\end{align}
	
	It is worth emphasizing that in the coordinate transformation, we only carried out it within the real number system. However, due to the Pauli matrices, it eventually presented as the complex cylindrical coordinate system $(\vec{ e}_{+1},\vec{ e}_{0},\vec{ e}_{-1})$, and the partial derivative operator $\vec{\nabla}$ has the form
	\begin{equation}
		\vec{ \nabla} = \sum_{\mu}(-1)^{\mu}\vec{ e}_{\mu}\nabla_{-\mu} = -\vec{ e}_{+1}\nabla_{-1} + \vec{ e}_{0}\nabla_{0} - \vec{ e}_{-1}\nabla_{+1},
		\label{eq:nabla_expansion}
	\end{equation}
	where 
	\begin{align}
		\nabla_{+1} &= -\frac{1}{\sqrt{2}}\left(\frac{\partial}{\partial x} + i\frac{\partial}{\partial y}\right) = -\frac{e^{i\theta}}{\sqrt{2}}\left(\frac{\partial}{\partial r} + \frac{i}{r}\frac{\partial}{\partial \theta}\right) \\
		\nabla_{-1} &= \frac{1}{\sqrt{2}}\left(\frac{\partial}{\partial x} - i\frac{\partial}{\partial y}\right) = \frac{e^{-i\theta}}{\sqrt{2}}\left(\frac{\partial}{\partial r} - \frac{i}{r}\frac{\partial}{\partial \theta}\right) \\
		\nabla_{0} &=
		\frac{\partial}{\partial z}
	\end{align}
	
	\section{Series Expansion Method}
	\label{sec: Expansion}
	
	Since the coefficients of each power of $r$ are zero we get four coupled recurrence relations:
	\begin{align}
		(\alpha + k - n )C_k^1 - \frac{ip_z}{\hbar}C_{k-1}^2 + \frac{i}{\hbar}\left(-\frac{E}{c} - mc\right)C_{k-1}^4 &= 0, \\
		(\alpha + k + n+1)C_k^2 + \frac{ip_z}{\hbar}C_{k-1}^1 + \frac{i}{\hbar}\left(-\frac{E}{c} - mc\right)C_{k-1}^3 &= 0, \\
		(\alpha + k - n)C_k^3 - \frac{ip_z}{\hbar}C_{k-1}^4 + \frac{i}{\hbar}\left(-\frac{E}{c} + mc\right)C_{k-1}^2 &= 0, \\
		(\alpha + k + n+1)C_k^4 + \frac{ip_z}{\hbar}C_{k-1}^3 + \frac{i}{\hbar}\left(-\frac{E}{c} + mc\right)C_{k-1}^1 &= 0
	\end{align}
	By simplifying the above recursive equations, we obtain the proportional relationships between the expansion coefficients in pairs:
	\begin{align}
		\frac{C_{k+1}^2}{C_k^1} &= -\frac{i}{\hbar}\frac{\lambda p_z + \left(-\frac{E}{c} - mc\right)}{\lambda(\alpha + k + n + 2)} , k \geq 0 \\
		\frac{C_{k+1}^4}{C_k^1} &= -\frac{i}{\hbar}\frac{p_z + \lambda \left(-\frac{E}{c} + mc\right)}{\lambda(\alpha + k + n + 2)}, k \geq 0 \\
		\frac{C_k^1}{C_{k-2}^1} &= \frac{C_k^3}{C_{k-2}^3} = \frac{1}{\hbar^2}\frac{-p_{\kappa}^2}{(\alpha + k + n)(\alpha + k - n)}, k \geq 2 \\
		\frac{C_{k+1}^2}{C_{k-1}^2}& = \frac{C_{k+1}^4}{C_{k-1}^4} = \frac{1}{\hbar^2}\frac{-p_{\kappa}^2}{(\alpha + k + n + 2)(\alpha + k - n)}, k \geq 1.
	\end{align}
	where we set that $p_{\kappa}^2 = \frac{E^2}{c^2} - m^2c^2 - p_z^2$ and $p_{\kappa}$ represents the transverse momentum. And the ratio of the two coefficients $C_k^1,C_k^3$ is a constant, that
	$\lambda=\frac{C_k^1}{C_{k}^3}$.
	To ensure the solution remains finite at the origin, the series must begin with non-negative powers. Thus, for $n \geq 0$, we set $\alpha = n$, and for $n<0$, we take $\alpha = -n-1$. They are the two roots of the indicial equation obtained by extracting the lowest power of $r$. So the final spinor wave function is also doubly degenerate. 
	
	Since the wave function obtained in both cases is degenerate, it does not affect the subsequent determination of eigenvalues or physical observables. Therefore, in the following we may focus on the case $n\geq 0, \alpha=n$. In the series expansion of $R_1,R_3$, all odd terms vanish, while in the series expansion of $R_2,R_4$, all even terms vanish. Substituting these parameters into the recursion relations and taking $k=2m$, we can obtain the expansion coefficient $C_{2m}^{1}$, 
	\begin{align}
		C_{2m}^1 = \frac{\kappa^{2m}(-1)^m}{2^{2m}m!n(n+1)(n+2)\cdots(m+n)} C_0
	\end{align}
	where $C_0$ is an arbitrary constant, typically chosen as $C_0 = \frac{1}{2^{n-1}\Gamma(n)}$, and $\kappa=p_{\kappa}/\hbar$ represents the quantum number of transverse momentum.
	Then the radial wave function $R_1(r)$ can be written as 
	\begin{equation}
		R_1(\kappa r) = C_0 \sum_{m=0}^{\infty} \frac{(-1)^m (\kappa r)^{2m+n}}{2^{2m}m!n(n+1)(n+2)\cdots(m+n)} 
	\end{equation}
	The power series solution for the radial function corresponds to Bessel functions which are defined as
	\begin{equation}
		J_{n}(\kappa r) = \sum_{m=0}^{\infty} \frac{(-1)^m}{m!\Gamma(m+n+1)}\left(\frac{\kappa r}{2}\right)^{2m+n}
	\end{equation}
	Once a solution for one component is obtained, the radial wave functions of the other spinor components can be derived using interdependent relationship between recursive formulas.

	\section{The Derivation of $\hat{K}$}
	\label{sec: Derivation}
	
	To derive this new conserved quantity $\hat{K}$, we consider the covariant form of the Dirac equation and  extend it from
	flat spacetime to curved spacetime by using Vierbeins \cite{Weyl} to transform into the curvilinear coordinate system and making a similarity transform into a frame where the Vierbeins are diagonal.
	
	In flat spacetime, the Dirac equation in cylindrical coordinates takes the form
	\begin{eqnarray}
		(\tilde{\gamma}^{\mu}\partial_{\mu}+mc^2)\psi=0
		\label{dirc2}
	\end{eqnarray}
	The connection between the Dirac matrices $\tilde{\gamma}^{\mu}$ and $\gamma^{\mu}$ is established by
%	\begin{widetext}
	\begin{eqnarray}
		\tilde{\gamma}^{0}=\gamma^{0}; \tilde{\gamma}^{1}=\gamma^{1}\cos \theta +\gamma^{2}\sin \theta; 
		\tilde{\gamma}^{3}=\gamma^{3}; \tilde{\gamma}^{2}=\frac{1}{r}(\gamma^{1}\sin \theta +\gamma^{2}\cos \theta);
	\end{eqnarray}
%	\end{widetext}
	where $\gamma^{\mu}$ is the gamma matrices in the Cartesian coordinate system, satisfying
	\[
	\gamma^0 = \begin{pmatrix} I & 0 \\ 0 & -I \end{pmatrix}, \quad
	\gamma^i = \begin{pmatrix} 0 & \sigma^i \\ -\sigma^i & 0 \end{pmatrix}
	(i=1,2,3)
	\]
	Then we introduce the similarity transformation, $S=e^{-\frac{\theta}{2}\gamma^1\gamma^2}$ as in the Ref \cite{Villalba1989}, such that 
	\begin{eqnarray}
		&&S^{-1}\tilde{\gamma}^{\mu}S=\bar{\gamma}^{\mu},\\
		&&S^{-1}\partial_{\mu}S=\partial_{\mu}-\Gamma_{\mu},\\
		&&S^{-1}\psi=\psi_d			
	\end{eqnarray}
	and the gamma matrices $\bar{\gamma}^{\mu}$ satisfy
	\begin{eqnarray}
		\bar{\gamma}^{0}=\gamma^{0}; \bar{\gamma}^{1}=\gamma^{1}; 
		\bar{\gamma}^{3}=\gamma^{3}; \bar{\gamma}^{2}=\frac{1}{r}\gamma^{2}.
	\end{eqnarray}
	The spin connection $\Gamma_{\mu}$ satisfies 
	\begin{eqnarray}
		\Gamma^{0}=\Gamma^{1}=\Gamma^{1}=0; \Gamma^{2}=-\frac{1}{2}\gamma^{1}\gamma^{2}.
	\end{eqnarray}
	The transformation $S$ acts on these matrices, reducing them to the rotating diagonal gauge. And $\psi_d$ is the diagonal Dirac wave function.
	In curved spacetime, the Dirac equation becomes
	\begin{eqnarray}
		[\gamma^{0}\partial_{t}+\gamma^{1}(\partial_{r}+\frac{1}{2r})+\gamma^{2}\frac{1}{r}\partial_{\theta}+\gamma^{3}\partial_{z}+mc^2]\psi_d=0	
		\label{curve-dirac}
	\end{eqnarray}
	To separate the variables, we let $\psi_d=\gamma^3\gamma^0\Phi$ and express Eq.(\ref{curve-dirac}) as the sum of two commute operators, $\hat{K_1}$ and $\hat{K_2}$, thereby isolating the time $t$, the mass $m$ and longitudinal $z$ dependence from the radial $r$ and polar angular $\theta$ dependence, and the corresponding eigenvalue equations can be written as
	\begin{eqnarray}
		&&\hat{K}_{1}\Phi
		=[\gamma^{1}(\partial_{r}+\frac{1}{2r})+\gamma^{2}\frac{1}{r}\partial_{\theta}]\gamma^3\gamma^0\Phi=\lambda'\Phi;\\
		&&\hat{K}_{2}\Phi
		=[\gamma^{0}\partial_{t}+\gamma^{3}\partial_{z}+mc^2]\gamma^3\gamma^0\Phi=-\lambda'\Phi
	\end{eqnarray}
	where $\pm \lambda'$ are the eigenvalues corresponding to the operators $\hat{K_1}$ and $\hat{K_2}$, respectively.
	The new conserved quantity for the vortex electron in cylindrical coordinates can be obtained from $\hat{K}_1$ with the help of the transformation $S$,
	\begin{eqnarray}
		S [\gamma^{1}(\partial_{r}+\frac{1}{2r})+\gamma^{2}\frac{1}{r}\partial_{\theta}]\gamma^3\gamma^0 S^{-1} S\Phi=\lambda' S\Phi 
		\label{k1s}
	\end{eqnarray}
	With $S \gamma^{\mu}S^{-1}=\gamma^{\mu}$, $S\partial_{\theta}S^{-1}=\partial_{\theta}$, $S\partial_{r}S^{-1}=0$, and $S\Phi=\tilde{\Phi}$, Eq. (\ref{k1s}) becomes
%	\begin{widetext}
	\begin{eqnarray}
		\hat{K}\tilde{\Phi}=[\gamma^{1}\gamma^{0}\gamma^{3}\cos \theta \frac{\partial}{\partial r}
		-\gamma^{1}\gamma^{0}\gamma^{3}\frac{\sin \theta}{r}\frac{\partial}{\partial \theta}
		+\gamma^{2}\gamma^{0}\gamma^{3}\sin \theta \frac{\partial}{\partial r}
		+\gamma^{2}\gamma^{0}\gamma^{3}\frac{\cos \theta}{r}\frac{\partial}{\partial \theta}]\tilde{\Phi}
		=\lambda'\tilde{\Phi} 
	\end{eqnarray}
%	\end{widetext}

\end{document}